\documentstyle[12pt,epsf]{article}
\newcommand{\beq}{\begin{equation}}
\newcommand{\eeq}{\end{equation}}
\newcommand{\ba}{\begin{array}}
\newcommand{\ea}{\end{array}}
\newcommand{\beqa}{\begin{eqnarray}}
\newcommand{\eeqa}{\end{eqnarray}}
\newcommand{\Frac}[2]{\frac{\displaystyle #1}{\displaystyle #2}}

\newcommand{\kpiee}{ $ K^+ \rightarrow \pi^+ e^+ e^- $   }
\newcommand{\kpiuu}{$K^+ \rightarrow \pi^+ \mu^+ \mu^- $ }
\newcommand{\kpll}{$K \rightarrow \pi \ell^+ \ell^-$ }

\newcommand{\opc}{{\cal O}(p^4)}
\newcommand{\ops}{{\cal O}(p^6)}
\newcommand{\gsim}{\stackrel{>}{_\sim}}
\newcommand{\lsim}{\stackrel{<}{_\sim}}

\renewcommand{\Im}{\mbox{Im~}}
\renewcommand{\Re}{\mbox{Re~}}

\begin{document}
\begin{titlepage}
\begin{flushright}
INFNNA-IV-98/25\\
UWThPh-1998-45\\
FTUV-98/61\\
IFIC-98/62\\
August 1998\\
\end{flushright}
\begin{center}

\vspace*{1cm}
 
{\large \bf The Decays $K \rightarrow \pi \ell^+ \ell^-$ \\[8pt]
beyond Leading Order in the Chiral Expansion*} 
 
\vspace*{1cm}
{\bf{G. D'Ambrosio$^1$, G. Ecker$^2$, G. Isidori$^3$ and J. Portol\'es$^4$}}
 
\vspace{.5cm}
${}^{1)}$ INFN, Sezione di Napoli \\
Dipartimento di Scienze Fisiche, Universit\`a di Napoli\\
I--80125 Napoli, Italy \\[5pt]
 
${}^{2)}$ Institut f\"ur Theoretische Physik, Universit\"at Wien\\
A--1090 Wien, Austria \\[5pt]
 
${}^{3)}$ INFN, Laboratori Nazionali di Frascati \\ 
P.O. Box 13, I--00044 Frascati, Italy\\[5pt]

${}^{4)}$ Departamento de F\'{\i}sica Te\'orica, Universidad de Valencia\\
E-46100 Burjassot, Spain\\[10pt]

{\bf Abstract} \\
\end{center}
We present a model--independent analysis of $K^+\to\pi^+\ell^+\ell^-$
and $K_S\to\pi^0\ell^+\ell^-$ decays, including $K\to 3 \pi$ unitarity 
corrections and a general decomposition of the dispersive amplitude.
 From the existing data on $K^+\to\pi^+ e^+ e^-$ we predict the 
ratio $R= B(K^+\to\pi^+\mu^+\mu^-)/B(K^+\to\pi^+ e^+ e^-)$ to be larger 
than 0.23, in slight disagreement with the recent measurement
$R = 0.167 \pm 0.036$.
Consequences for the $K^\pm \to\pi^\pm e^+ e^-$ charge asymmetries
and for the $K_L\to\pi^0 e^+ e^-$ mode are also discussed. 
\noindent

\vfill
\noindent * Work supported in part
by TMR, EC--Contract No. ERBFMRX--CT980169 \\(EURODA$\Phi$NE)

\end{titlepage}

\section{Introduction}

Radiative nonleptonic kaon decays continue to provide interesting
information on the structure of the weak interactions at low energies.
Among them, the flavour--changing neutral current (FCNC)  transitions
$K \rightarrow \pi \ell^+ \ell^-$,  
induced at the one--loop level in the Standard
Model, are well suited to explore its 
quantum structure and, possibly, its extensions \cite{EPR1,DEIN}.
On the experimental side, the high--precision measurements already 
achieved at Brookhaven (AGS) \cite{AGS98}
and  Fermilab (KTeV) \cite{KTEV98}, and foreseen in the near
future at Frascati (KLOE) \cite{DPHB}, call for a thorough
theoretical investigation.

The $K^+ \rightarrow \pi^+\ell^+ \ell^-$ and $K_S \rightarrow 
\pi^0 \ell^+ \ell^-$ ($\ell = e, \mu$) channels, expected to be dominated
by long--distance dynamics through one--photon exchange
($K \rightarrow \pi \gamma^*$),  were studied at leading order in the
chiral expansion in Ref.~\cite{EPR1}. Once the unknown
combination of local terms arising in this framework is fixed by
comparison with $K^+ \rightarrow \pi^+ e^+ e^-$ data \cite{AL92},
the process $K^+ \rightarrow \pi^+ \mu^+ \mu^-$ can be predicted
without further assumptions\footnote{The neutral channels (still
unmeasured) require additional model--dependent assumptions as
discussed in Sect.~\ref{sect:insights}.}. Recently, this last
mode has been observed \cite{AD97}, yielding a value for 
$R=\Gamma(K^+ \rightarrow \pi^+ \mu^+ \mu^-) / \Gamma(K^+ \rightarrow \pi^+
 e^+ e^-)$ that is $\sim 2 \, \sigma$ below the leading--order 
prediction. This experimental result has
motivated the present study of $K \rightarrow \pi \gamma^*$ form factors 
beyond the leading order in
the chiral expansion. We have carried out this program by including
unitarity corrections from $K \rightarrow \pi \pi \pi$ together with
the most general polynomial structure consistent with the 
chiral expansion up to $\ops$. In this way we perform a
model--independent analysis of the existing
$K^+ \rightarrow \pi^+ \ell^+ \ell^- $ data leading to
a consistent fit of both rate and dilepton invariant--mass spectrum 
of the electron channel. On the other hand,
we demonstrate that the persisting discrepancy in the ratio $R$
cannot be accommodated in the Standard Model.

Besides this phenomenological analysis, we 
present a general discussion of theoretical 
predictions for the polynomial part 
of the $K^+(K_S) \rightarrow \pi^+(\pi^0) \gamma^*$ form factors.
We show that it is very difficult at present to 
estimate  the branching ratios of the 
$K_S \rightarrow \pi^0 \ell^+ \ell^-$ modes
without strong model--dependent assumptions. 
More accessible are the  dilepton invariant--mass spectrum 
and, correspondingly, the ratio 
$\Gamma(K_S \rightarrow \pi^0 \mu^+ \mu^-) / \Gamma(K_S \rightarrow \pi^0
 e^+ e^-)$.

We conclude our analysis by discussing the impact
of the long--distance $K\to \pi\gamma^*\to \pi e^+ e^-$
transitions for estimating $CP$--violating observables.
In particular, we analyse the possibilities for
disentangling direct and indirect $CP$--violating 
components of the $K_L \rightarrow \pi^0 e^+e^-$ amplitude.
Moreover, we estimate the effect of the unitarity corrections 
on the charge asymmetry of $K^{\pm} \rightarrow \pi^{\pm} e^+ e^-$
decays.

The plan of the paper is the following:
in the next section we present the model--independent analysis of
$K \rightarrow \pi \gamma^*\to \pi \ell^+\ell^-$ transitions.
Section \ref{sect:insights} is devoted to explore the physics 
behind the polynomial coefficients of the $K \rightarrow \pi \gamma^*$
form factors. We turn to a discussion of  $CP$--violating
observables in Sect.~\ref{sect:CPV}. 
Our main conclusions are summarized in the last section.

\section{Model--independent analysis}
\label{sect:MIA}

\paragraph{2.1} The FCNC transitions  \kpll ($\ell=e,\mu$) 
are dominated by single virtual photon exchange 
($K\to \pi\gamma^*\to\pi \ell^+ \ell^-$) if allowed
by $CP$ symmetry. This is the case for $K^+$ and $K_S$ decays
where the amplitude is  determined by an electromagnetic
transition form factor in the presence of the nonleptonic weak
interactions:
\beq
i \int d^4x e^{iqx} \langle \pi(p)|T \left\{J^\mu_{\rm elm}(x)
{\cal L}_{\Delta S=1}(0) \right\} |
K(k)\rangle = \Frac{W(z)}{(4\pi)^2}\left[z(k+p)^\mu -(1-r_\pi^2)q^\mu
\right]~,
\label{eq:tff}
\eeq
$$
k^2=M_K^2~,\quad p^2=M_\pi^2~,\quad q=k-p~,\quad 
z=q^2/M_K^2~,\quad r_\pi = M_\pi /M_K~,
$$
where ${\cal L}_{\Delta S=1}$ is the strangeness changing
nonleptonic weak Lagrangian and $J^\mu_{\rm elm}$ is the
electromagnetic current.
The dynamics of the decays is completely
specified by the invariant functions $W_+(z)$ and $W_S(z)$. 
As dictated by gauge invariance, these functions vanish to
lowest order in the low--energy expansion \cite{EPR1}. To account
for this chiral suppression we have pulled out a
factor $1/(4\pi)^2$ in the definition (\ref{eq:tff}).
Since the Cabibbo angle is approximately compensated by the
nonleptonic octet enhancement, the natural magnitude of $W(z)$ is 
expected to be $G_F M_K^2$. 

With these conventions\footnote{In Ref.~\cite{DEIN} the 
form factor $V(z)=W(z)/G_8 M_K^2$  was used instead of $W(z)$.
Here we prefer to avoid introducing
the coupling constant $G_8$ in the model--independent analysis.
We use the occasion to point out two misprints in
Eq.~(4.27) of \cite{DEIN}; the correct formula is given by 
(\ref{fspectrum}).} the decay amplitude takes the form
\beq
A\left( K(k) \rightarrow \pi(p) \ell^+(p_+) \ell^-(p_-) \right)  
=  - \Frac{e^2}{M_K^2 (4 \pi)^2} W(z)
(k+p)^\mu \bar{u}_\ell(p_-) \gamma_\mu v_\ell(p_+)~.
\eeq
The spectrum in the dilepton invariant mass is then given by 
\begin{equation}
\Frac{d\Gamma}{d z} \, = \, \Frac{\alpha^2 M_K}{12 \pi (4
\pi)^4} \,
\lambda^{3/2}(1,z,r_{\pi}^2) \, \sqrt{1 - 4 \Frac{r_{\ell}^2}{z}} \,
\left( 1 + 2 \Frac{r_{\ell}^2}{z} \right) \, | W(z) |^2 \; ,
\label{fspectrum}
\end{equation}
with $r_\ell = m_\ell/M_K$ and $ 4 r_{\ell}^2 \leq z \leq (1-r_{\pi})^2$.

\paragraph{2.2} The form factors $W_i(z)$ ($i=+,S$) are
analytic functions in the complex $z$--plane cut along the positive
real axis. The cut starts at $z=4 r_\pi^2$ with the two--pion
threshold. For the small dilepton masses occurring in the decays,
one expects the $\pi^+\pi^-$ intermediate state to play the dominant role
in the dispersion relations for $W_i(z)$. The
contribution of higher--mass intermediate states can be described
by a low--order polynomial in $z$. This is
known to be a very good approximation for the $K^+K^-$ contribution,
for instance \cite{EPR1}.

\begin{figure}[t]
    \begin{center}
       \setlength{\unitlength}{1truecm}
       \begin{picture}(8.0,4.0)
       \epsfxsize 8.0 true cm
       \epsfysize 4.0 true cm
       \epsffile{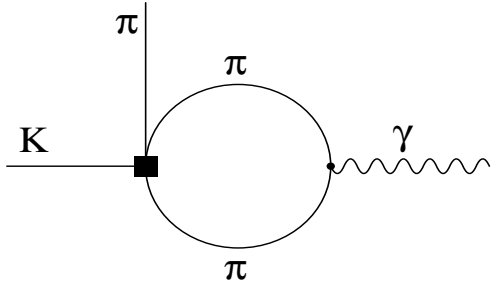}
       \end{picture} 
    \end{center}
    \caption{$K\to 3\pi$ contribution to the effective 
$K\to\pi\gamma^*$ vertex.}
    \protect\label{fig:loop}
\end{figure}

We therefore decompose the form factors $W_i(z)$ as
\beq
W_i(z)= G_F M_K^2 W_i^{\rm pol}(z)+ W_i^{\pi\pi}(z)~,
\label{ffdecomp}
\eeq
where $W_i^{\pi\pi}(z)$ denotes the contribution from the two--pion 
intermediate state. To leading nontrivial order in the chiral
expansion \cite{EPR1}, the $W_i^{\rm pol}(z)$ are constants (but for 
a negligible contribution due to the kaon loop). The
functions $W_i^{\pi\pi}(z)$ can be calculated from the diagram shown in 
Fig.~\ref{fig:loop}, where the $K\to 3 \pi$ vertex is taken from the
leading nonleptonic weak Lagrangian of ${\cal O}(p^2)$.

Here we go one step further. We use the physical $K\to3\pi$ amplitude 
expanded up to $\opc$ (as required by present 
$K\to3\pi$ data) and include the electromagnetic form factor $F(z)$
(normalized to $F(0)=1$) for $\pi^+\pi^-\to\gamma^*$ to first
nontrivial order. The relevant  $K\to3\pi$ amplitudes are expanded as
\cite{DIPP}
\beqa
A(K^+ \to \pi^+ \pi^+ \pi^- ) &=& 2a_c + (b_c + b_2) Y + 2 c_c
(Y^2+X^2/3) 
+(d_c+d_2) (Y^2-X^2/3)~, \nonumber\\ 
A(K_S \to \pi^+ \pi^- \pi^0 ) &=& \frac{2}{3} b_2 X - \frac{4}{3}d_2
XY~, 
\label{K3pi}
\end{eqnarray}
with
\beq
s_i= (k-p_i)^2~,\qquad 
s_0= {1\over 3} (s_1+s_2+s_3)~,\qquad 
X= { s_1 - s_2 \over M_\pi^2 }~,\qquad 
Y= { s_3 - s_0 \over M_\pi^2 }~, 
\eeq
where $p_i$ denote the pion momenta\footnote{The subscript 3 indicates
the odd--charge pion, i.e. $\pi^-$ in $K^+ \to \pi^+ \pi^+ \pi^-$ and
$\pi^0$ in $K_S \to \pi^+ \pi^- \pi^0$.}. To the same level of
accuracy, the electromagnetic  form factor is  
$F(z) = 1+ z/r^{2}_V$ with $r^2_V = M_V^2/M_K^2 \simeq 2.5$. In accordance
with chiral counting, the polynomial in (\ref{ffdecomp}) is
assumed to have the general form
\beq
W_i^{\rm pol}(z)= a_i +  b_i z \qquad (i=+,S) ~.
\label{eq:Wpol}
\end{equation}

Up to a linear polynomial of this type, the dispersion integral
corresponding to Fig.~\ref{fig:loop} is unambiguously calculable with the
result 
\beq
W_i^{\pi\pi}(z) = {1\over r_\pi^2}~ 
\left[\alpha_i + \beta_i \frac{z-z_0}{r_\pi^2}\right]~F(z)~\chi(z)~,
\label{eq:Wpp}
\eeq
where $z_0=1/3+r^2_\pi$. The one--loop function
\beq
\chi(z) = {4\over 9} - {4r_\pi^2\over 3 z} - {1\over 3}(1 - 
{4r_\pi^2\over z}) G(z/r_\pi^2)
\eeq
$$
 G(z) = \Biggl\{
   \ba{ll}
      \sqrt{4/z -1} \arcsin{(\sqrt{z}/2)} \qquad & z\le 4 \\
      -{1\over 2}\sqrt{1-4/z}
      \left( \ln\Frac{1-\sqrt{1-4/z}}{1+\sqrt{1-4/z}} + i\pi \right)
       & z\ge 4
   \Biggr. \ea 
$$
satisfies $\chi(0)=0$. 
In terms of the $K\to3\pi$ parameters in (\ref{K3pi}), we find
\beq
\alpha_+ =    -(b_c+b_2)~, \qquad
\beta_+ =  2(d_c+d_2)~, \qquad
\alpha_S =   \Frac{4}{3} b_2~, \qquad
\beta_S =   - \Frac{8}{3} d_2~.
\label{eq:cd}
\eeq

The total form factor, given by 
\beq
W_i(z)=G_F M_K^2 (a_i + b_i z) + W_i^{\pi\pi}(z)~,
\label{eq:Wtot}
\eeq
is expected to be an excellent approximation to the complete form factor
of $\ops$. The polynomial piece has the most general form that can
occur to this order, with a priori unknown low--energy constants
contributing to the $a_i$, $b_i$. The main assumption underlying the
form factor (\ref{eq:Wtot}) is that all other contributions to the dispersion
integral except the two--pion intermediate state can be well
approximated by a linear polynomial for small values of $z$.

\begin{table}[t]
\begin{center}
\begin{tabular}{|c|c|c|c|} 
\hline
$b_c$          & $b_2$          & $d_c$           & $d_2$          \\ \hline
24.5 $\pm$ 0.3 & -3.9 $\pm$ 0.4 & -1.6 $\pm$ 0.3  & 0.2 $\pm$ 0.5  \\ \hline
\end{tabular}
\caption{Experimental values of $K\to 3\pi$ amplitudes
\protect\cite{KMW2} contributing to
the $W_i^{\pi\pi}(z)$. All entries are in units of $10^{-8}$.}
\label{tab:bi}
\end{center}
\end{table}

Using the central values of the $K\to3\pi$ parameters \cite{KMW2}
in Table \ref{tab:bi}, 
we can now calculate the branching ratios for 
$K^+ \rightarrow \pi^+ \ell^+ \ell^-$ and
$K_S \rightarrow \pi^0 \ell^+ \ell^-$ 
in terms of the corresponding parameters:
\begin{eqnarray}
B(K^+ \rightarrow \pi^+ e^+ e^-) \! &=& \!
\left[ 0.14 - 3.23 a_+ - 0.88 b_+ + 59.2 a_+^2 + 16.0 a_+ b_+ + 1.73
b_+^2 \right]\times 10^{-8}~, \nonumber \\
B(K^+ \rightarrow \pi^+ \mu^+ \mu^-) \! &=& \!
\left[ 1.13 - 19.2 a_+ - 6.32 b_+  + 116 a_+^2  + 67.3 a_+ b_+ + 10.3
b_+^2 \right]\times 10^{-9}~, \nonumber \\
B(K_S \rightarrow \pi^0 e^+ e^-) \! &=& \!
\left[ 0.01 - 0.76 a_S - 0.21 b_S + 46.5 a_S^2 + 12.9 a_S b_S + 1.44
b_S^2 \right]\times 10^{-10}~, \nonumber \\
B(K_S \rightarrow \pi^0 \mu^+ \mu^-) \! &=& \!
\left[ 0.07 - 4.52 a_S - 1.50 b_S + 98.7 a_S^2 + 57.7 a_S b_S + 8.95
b_S^2 \right]\times 10^{-11}~. \nonumber \\
\label{eq:BRlist}
\end{eqnarray}

With $a_+$ and $a_S$ expected to be of ${\cal O}(1)$, we observe that 
the amplitudes and branching ratios are actually dominated by the
polynomial parts. $W_i^{\pi\pi}(z)$ in (\ref{eq:Wpp}) contributes
to the rate mainly through its interference with 
(\ref{eq:Wpol}). In fact, this really only applies to the $K^+$ mode
because the two--pion contribution is practically negligible for the
neutral decay, due to the strong suppression of the $K_S\to \pi^+
\pi^- \pi^0$ amplitude.

\paragraph{2.3} In the Brookhaven experiment
BNL-E777 \cite{AL92} both spectrum and branching ratio of the decay
$K^+ \rightarrow \pi^+ e^+ e^-$  were measured.
They fit a spectrum with two parameters $C$ and $\lambda$
defined through
\begin{equation}
\Frac{d \Gamma}{d M_{ee}} \, = \, \Frac{C}{8} \, M_{ee} \, M_K^3 \, 
\lambda^{3/2}(1,r_{\pi}^2,\Frac{M_{ee}^2}{M_K^2}) \, 
\left( 1 \, + \, \lambda \, \Frac{M_{ee}^2}{M_{\pi}^2} \right)^2 \, ,
\label{eq:spectrum}
\end{equation}
where $M_{ee}$ is the invariant mass of the lepton pair ($M_{ee} = 
M_K \sqrt{z}$), obtaining
\begin{eqnarray}
\lambda \, & = & \, 0.105 \, \pm \, 0.035 \, \pm \, 0.015~,
\label{eq:lambda} \\
B(K^+ \rightarrow \pi^+ e^+ e^-) \, & = & \, (2.75 \, \pm \, 0.23
\, \pm 0.13 ) \times 10^{-7}~. \label{eq:BRtt}
\end{eqnarray}
However, in order to subtract the background due to the process
$K^+ \rightarrow \pi^+ \pi^0$, $\pi^0 \rightarrow e^+ e^- \gamma$, they
make a cut in the dilepton invariant mass and consider only 
$K^+ \rightarrow \pi^+ e^+ e^-$ events with $M_{ee} > 0.150$~GeV. 
Thus the branching ratio actually measured is given by\footnote{We have
extracted this value from (\ref{eq:lambda}) and (\ref{eq:BRtt}) using
(\ref{eq:spectrum}) for the spectrum. For a  conservative
estimate of the error, we have scaled down the error given in 
(\ref{eq:BRtt}).}

\begin{equation}
B(K^+ \rightarrow \pi^+ e^+ e^-) \, |_{cut} \, = \, (1.81 \, \pm \,
0.17)\times 10^{-7}
\label{eq:BRcut}
\end{equation}  
and the result in (\ref{eq:BRtt}) includes a theoretical
extrapolation to the low $M_{ee}$ region.
            
\begin{figure}[t]
    \begin{center}
       \setlength{\unitlength}{1truecm}
       \begin{picture}(10.0,10.0)
       \epsfxsize 10.0 true cm
       \epsfysize 10.0 true cm
       \epsffile{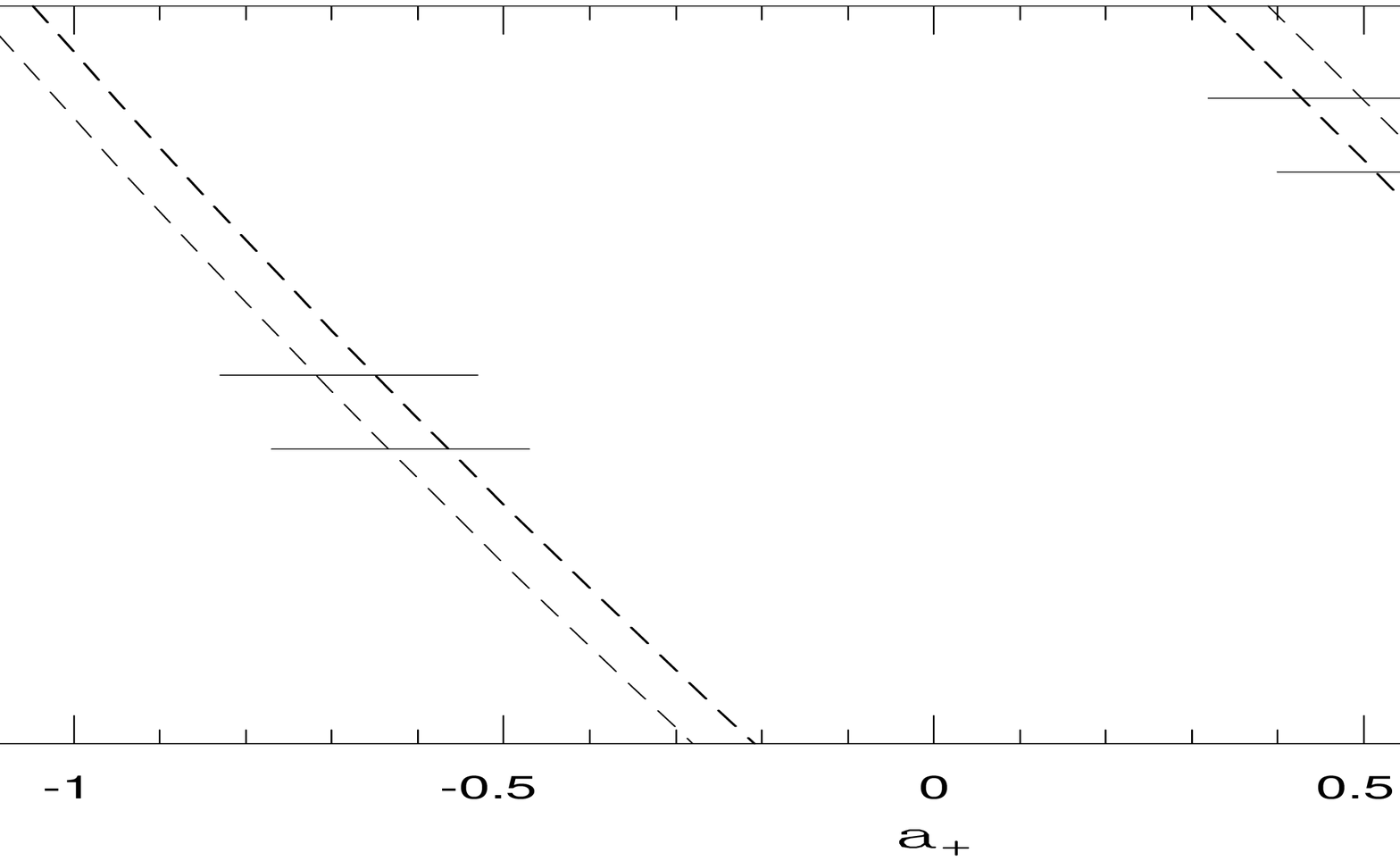}
       \end{picture} 
    \end{center}
    \caption{Allowed domains in the $a_+$--$b_+$ plane (inside the
       dashed curves) from the experimental branching ratio 
       (\protect\ref{eq:BRcut}). The bounds on $b_+$ from the measured
       slope (\protect\ref{eq:lambda}) are also shown (full lines).}
    \protect\label{fig:abdomain}
\end{figure}

The branching ratio (\ref{eq:BRcut}) can be translated into allowed
domains for the parameters $a_+$ and $b_+$ as shown in
Fig.~\ref{fig:abdomain}. For a generous range $|b_+|\le 2$, two branches
of solutions are in principle possible with opposite signs of
$a_+$.

\begin{figure}[t]
    \begin{center}
       \setlength{\unitlength}{1truecm}
       \begin{picture}(10.0,10.0)
       \epsfxsize 10.0 true cm
       \epsfysize 10.0 true cm
       \epsffile{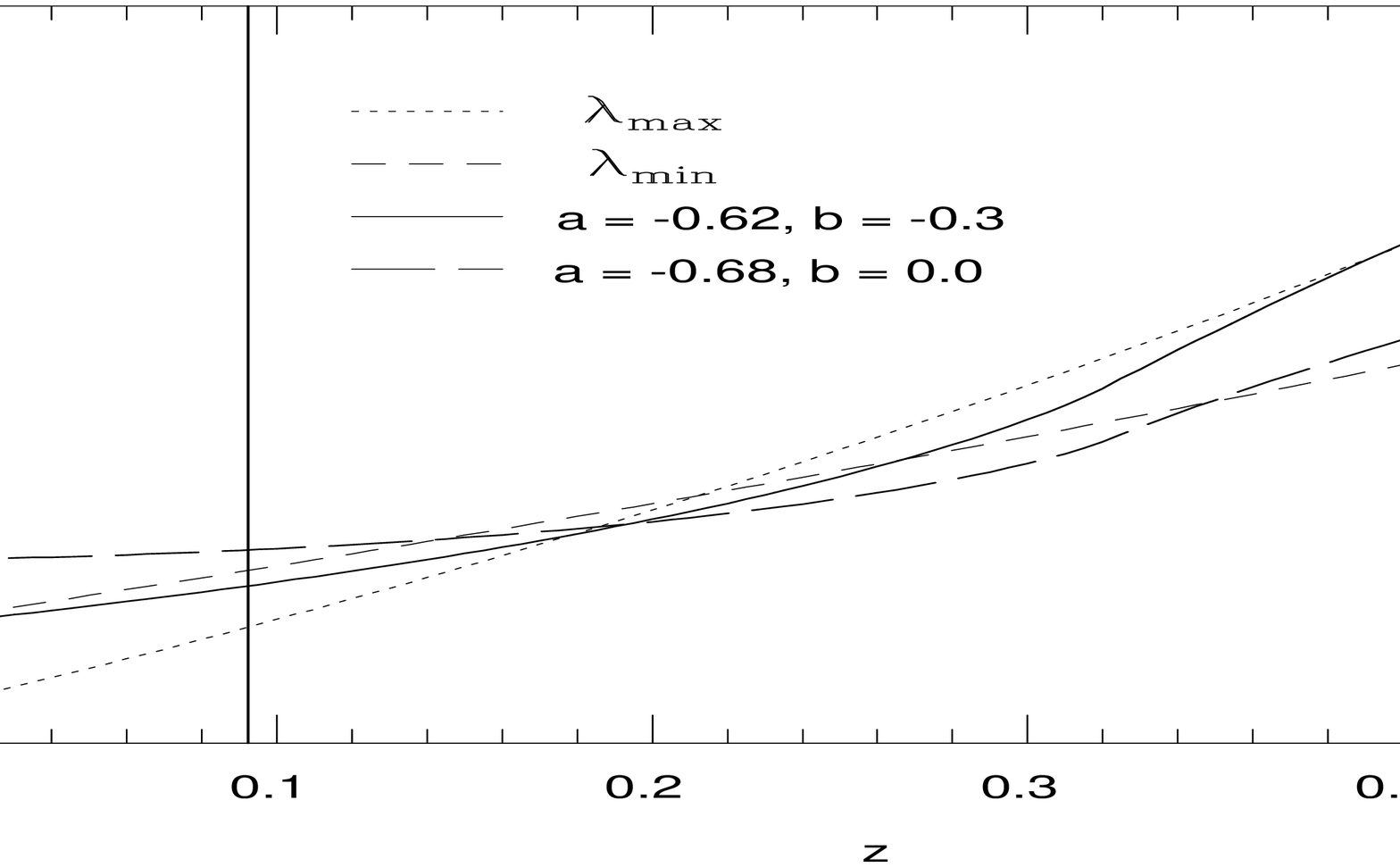}
       \end{picture} 
    \end{center}
    \caption{Comparison between different shapes of the \kpiee form 
factor for $a_+<0$. The dotted and short--dashed lines correspond to
the $\pm 1 \sigma$ values of the $\lambda$ parameter  
in (\protect\ref{eq:lambda}). The solid and long--dashed lines 
correspond to the theoretical prediction for two sets of values
of $a_+$ and $b_+$. The comparison is to be applied 
for $M_{ee} >  0.150 \, \mbox{GeV}$ (vertical line) as measured by 
experiment \protect\cite{AL92}. }
    \protect\label{fig:Wplot1}
\end{figure}

Unlike the rate, the spectrum is very sensitive to $b_+$. The
bounds on $b_+$ from the experimental spectrum are also 
indicated in Fig.~\ref{fig:abdomain}. Let us try to understand the
allowed values of $a_+$ and $b_+$ with the help of the amplitude
(\ref{eq:Wtot}). The real part of the two--pion contribution, 
$\Re W_+^{\pi\pi}(z)$, is a monotonically
decreasing function of $z$ in the physical region. Assuming first
$b_+=0$, as would be the case at $\opc$ (except for the tiny
contribution from the kaon loop), we infer that $a_+$ must be
negative to reproduce the experimentally observed positive slope 
(\ref{eq:lambda}). In fact, the spectrum prefers a somewhat bigger
slope requiring $b_+/a_+ >0$. This is made more explicit in 
Fig.~\ref{fig:Wplot1}, where the theoretical spectrum (or rather the 
square of the form factor) is confronted with the experimental slope. 

\begin{figure}[t]
    \begin{center}
       \setlength{\unitlength}{1truecm}
       \begin{picture}(10.0,10.0)
       \epsfxsize 10.0 true cm
       \epsfysize 10.0 true cm
       \epsffile{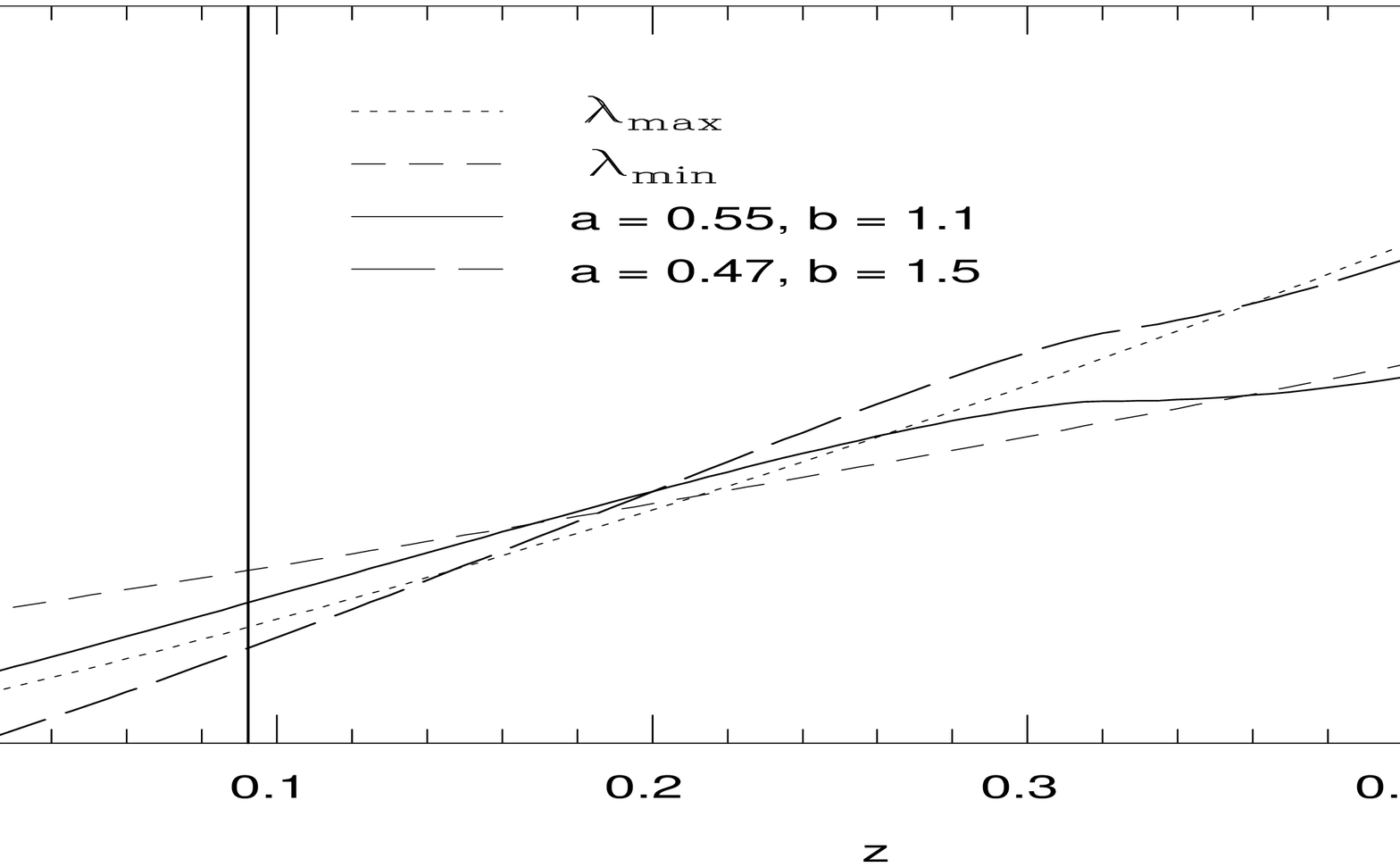}
       \end{picture} 
    \end{center}
    \caption{Comparison between different shapes of the \kpiee form 
factor for $a_+>0$ (as in
       Fig.~\protect\ref{fig:Wplot1}).} 
    \protect\label{fig:Wplot2}
\end{figure}

At $\opc$ only negative values of $a_+$ are allowed \cite{AL92}. 
However, in general there can be a second branch of
solutions with positive $a_+$. Since in this case the slope is negative 
for $b_+=0$, a sufficiently large $b_+$ ($b_+/a_+ >1$) is necessary to 
reproduce the observed spectrum. Analogously to Fig.~\ref{fig:Wplot1},
we exhibit the square of the form factor in
Fig.~\ref{fig:Wplot2} for two sets of $a_+$ and $b_+$ corresponding to
the $\pm 1\sigma$ values of the measured slope.

As we shall discuss in the next section, 
the second branch of solutions with
positive $a_+$ is disfavoured from a theoretical point of view.
Indeed, naive chiral counting would suggest that
the coefficient $b_+$, which only arises at $\ops$, 
should be smaller in magnitude than the
leading coefficient $a_+$.

For the present analysis we do not discard the second type
of solutions right away but confront the theoretical amplitude with
another piece of experimental information that has recently become
available through the measurement of the muonic decay mode of
$K^+$. The Brookhaven experiment of Adler et al. \cite{AD97} has
measured a branching ratio
\beq
B(K^+ \rightarrow \pi^+ \mu^+ \mu^-)  =  (5.0 \pm 0.4 ({\rm stat}) \pm
0.7 ({\rm syst}) \pm 0.6 ({\rm th}) ) \times 10^{-8}
\eeq
which, together with the results from \cite{AL92}, implies a muon/electron
ratio
\beq
R \, = \, \Frac{B(K^+ \rightarrow \pi^+ \mu^+ \mu^-)}{B(K^+
\rightarrow \pi^+ e^+ e^-)} \, = \, 0.167 \pm 0.036 ~.
\label{eq:Rexp}
\end{equation}

Can we use this experimental input to distinguish between the two
branches in the $a_+-b_+$ plane? Looking first at
the solutions with $a_+<0$, we find that $R$ is a
monotonically increasing function of $|b_+|$ for all allowed values
in the left branch of Fig.~\ref{fig:abdomain} and it
is therefore always bigger than 0.23 (obtained for $b_+=0$). 
At first sight the second branch looks more promising for
understanding the experimental result (\ref{eq:Rexp}). In fact, for
$b_+=0$ the ratio $R$ is small enough \cite{EPR1} to agree
with (\ref{eq:Rexp}). However, as Fig.~\ref{fig:abdomain} shows, the
experimental spectrum for the electronic decay channel
requires $b_+>a_+$. Since $R$ is an increasing function of $b_+$ for
$a_+>0$, the resulting values of $R$ turn out to be even larger than
in the previous case. Therefore, the theoretically unattractive
solutions with positive $a_+$ cannot explain the small $\mu/e$ ratio
either. For the values of $a_+$, $b_+$ used in 
Figs.~\ref{fig:Wplot1},~\ref{fig:Wplot2} the results are collected in 
Table \ref{tab:Rmue}.

\begin{table}
\begin{center}
\begin{tabular}{|r|r|c|r|} 
\hline
\multicolumn{1}{|c|}{$a_+$} & 
\multicolumn{1}{|c|}{$b_+$} & 
\multicolumn{1}{|c|}{$10^8$ B(\kpiuu)} &  
\multicolumn{1}{|c|}{$R$}  \\
\hline
\hline
$-$0.68 & 0.0 & 6.78 & 0.228 \\
\hline
$-$0.62 & $-$0.3 & 7.29 & 0.258 \\
\hline
0.55 & 1.1 & 7.19 & 0.265 \\
\hline
0.47 & 1.5 & 7.89 & 0.309 \\
\hline
\end{tabular} 
\caption{$B(K^+ \rightarrow \pi^+ \mu^+ \mu^-)$  and the ratio $R$ as a
function of $a_+$ and $b_+$.}
\label{tab:Rmue}
\end{center}
\end{table}

On the basis of this model--independent analysis, we conclude that the
central values of the $\mu/e$ ratio  \cite{AD97} and the
slope of the electron spectrum \cite{AL92} together are not 
consistent. The $\sim 2 \sigma$ discrepancy can be due to a statistical
fluctuation or could be an indication of peculiar non--standard physics. 
We stress that the inconsistency is essentially unrelated to 
the chiral expansion but only depends on the assumption that both channels
are dominated by the same $K\to\pi\gamma^*$ form factor.
Indeed, a form factor $W_+(z)$ rising with $z$ as indicated by   
$K^+\to\pi^+ e^+ e^-$ data implies $R> R_{\rm phase-space}=0.196$ \cite{EPR1}.

\section{Theoretical ideas on $a_i$ and $b_i$}
\label{sect:insights}

The leading $\opc$ predictions of $a_i$ and $b_i$ are given by 
\beqa
a_+^{(4)} &=& \Frac{G_8}{G_F} \left( 1/3 - w_+ \right)~, \nonumber   \\
a_S^{(4)} &=& - \, \Frac{G_8}{G_F} \left( 1/3 - w_S \right)~,  \nonumber  \\
b_+^{(4)} &=& - \, \Frac{G_8}{G_F} \, \Frac{1}{60}~,  \nonumber       \\
b_S^{(4)} &=& \, \Frac{G_8}{G_F} \, \Frac{1}{30}~,
\label{eq:asbs}
\eeqa
in terms of the usual  parameters $G_8$ and $w_i$ \cite{EPR1}. 
Local contributions to the $b_i$ are forbidden at $\opc$
and the small values in (\ref{eq:asbs}) are generated 
by the expansion of the kaon--loop function. Sizable corrections
to the $b_i$ are expected at $\ops$ where  
local terms are  allowed. However, since local terms 
contribute to the $a_i$ already at $\opc$ ($w_i$ terms), 
naive chiral counting would suggest $b_i/a_i \sim {\cal O}(p^6)/{\cal O}(p^4) < 1$.
The solution with both $a_+$ and $b_+$ negative
discussed in  the previous section satisfies
this expectation.
\par
The expressions of $w_+$ and $w_S$ in terms of the 
$\opc$ low--energy couplings $N_i$ \cite{EKW93}, $L_9$ \cite{GL85} are
\begin{eqnarray}
w_{+} \, & = & \, \Frac{64 \pi^2}{3} \, \left[ \, N_{14}^r(\mu) \, - \, 
N_{15}^r(\mu) \, + \, 3 L_9^r(\mu) \, \right] \, + \, \Frac{1}{3} \ln \left( 
\Frac{\mu^2}{M_K M_{\pi}} \right)~, \nonumber \\
w_S \, & = & \, \Frac{32 \pi^2}{3} \, \left[ \, 2 N_{14}^r(\mu) \, + 
\, N_{15}^r(\mu) \, \right] \, + \, \Frac{1}{3} \ln \left(
\Frac{\mu^2}{M_K^2} 
\right)~. \label{eq:kpgct}
\end{eqnarray}
Unfortunately, the 
values of the  $N_i^r$ are not known and to make definite predictions 
we need to rely on model--dependent assumptions. For this purpose,
it is interesting to note that while the combinations of
low--energy constants appearing 
in $w_+$ and $w_S$ (or $a_+$ and $a_S$)
depend separately on the renormalization scale $\mu$, 
the combination occurring in $a_+ + a_S$ does not. 
As originally noted in   \cite{EPR1},
this scale independence might be related to 
the structure of the effective four--fermion Hamiltonian relevant for 
$K \rightarrow \pi \ell^+ \ell^-$. Indeed,
as we will discuss in the next section, there is only one 
dimension--six operator giving a non--vanishing contribution to
the $a_i$ at leading order:
$Q_{7V}  =  \overline{s}\gamma^\mu(1-\gamma_5) d \overline{\ell} 
\gamma_{\mu} \ell$.
Due to the octet structure of the $\overline{s} \gamma^{\mu} d$ current, 
this operator does not affect the combination $a_+  +  a_S$. 
Actually, the cancellation of  this  short--distance
contribution holds not only for   $a_+  +  a_S$ but (in the limit of 
isospin conservation) for $A(K^+\to\pi^+\ell^+\ell^-)$ $+$ 
$A(K_S\to\pi^0\ell^+\ell^-)$ in general, 
that would thus be completely determined by low--energy
dynamics. As a consequence, we find it more reliable to predict the
sum rather than the separate expressions of $W_{+}^{\rm pol}$ and 
$W_S^{\rm pol}$ using a low--energy model. In the following we shall 
employ the Vector Meson Dominance (VMD) hypothesis 
to estimate  $W_+^{\rm pol} +  W_S^{\rm pol}$.
  
The vector meson contributions to the low--energy constants 
of the $\opc$ weak Lagrangian have been discussed in \cite{EKW93,DP97}.
Employing the vector field formulation of Ref.~\cite{DP97},
the result for $W_+^{\rm pol} +  W_S^{\rm pol}$
is independent of the factorization hypothesis
and can be specified in terms of a single unknown parameter $\eta_V$
\begin{equation}
W_{+,V}^{(4)} \,  +  \, W_{S,V}^{(4)} = \, \Frac{G_8}{G_F} \, 
\left[ \, 16 \pi^2 \, f_V^2 \, ( \, 2 \eta_V \, - \, 1 \, ) \, + \, 
\Frac{1}{3} \, \ln \left( \Frac{M_{\pi}}{M_K} \right) \, \right]~,  
 \label{eq:w4}
\end{equation}
where $|f_V| \simeq 0.20$, as obtained from 
$\Gamma(\rho^0 \rightarrow e^+ e^-)$. The logarithmic 
term in (\ref{eq:w4}) is a residual effect of the loop 
amplitudes, whereas the term proportional to  
$f_V$ is the local vector meson contribution.
The $\eta_V$ parameter 
is not known but is expected to be in the range
$0 \lsim \eta_V  \lsim 1$; in principle it 
is measurable in other processes \cite{DP97}.
Note that for typical values of $\eta_V$
(but for $\eta_V\simeq 1/2$) the local vector meson contribution is dominant.
A relation equivalent to (\ref{eq:w4}) was obtained in \cite{EKW93}
under the factorization assumption (the result of \cite{EKW93}
is obtained from  (\ref{eq:w4}) with the substitution $\eta_V\to k_F$).
As discussed in \cite{EKW93}, the separate 
predictions of $a_S$ and $a_+$ involve an additional unknown 
coupling which is difficult to interpret in the framework of VMD.

The evaluation of the $\ops$ local terms generated by
vector meson exchange is more involved \cite{DP96,DP98}.
In general, we can identify  two kinds of contributions:
those originating from genuine $\ops$ weak transitions
and those generated by the pole expansion of the leading $\opc$ 
VMD results (we refer to \cite{DP96,DP98} for a 
detailed discussion). Interestingly enough, all the effects of the 
genuine $\ops$ weak transitions, evaluated under the factorization assumption,
drop out in the sum $W_+^{\rm pol} + W_S^{\rm pol}$. Moreover, the $\ops$ 
vector exchange 
contributions only modify the slope parameters $b_i$, leading to
the simple result
\begin{equation}
W^{(6)}_{+,V} + W^{(6)}_{S,V}  =  \, \Frac{G_8}{G_F} \, {z\over r_V^2}\, 
\left[ \, 16 \pi^2 \, f_V^2 \, ( \, 2 \eta_V \, - \, 1 \, ) \, \right]~. 
 \label{eq:w6}
\end{equation}
Combining (\ref{eq:w4}) and (\ref{eq:w6}) and neglecting the 
presumably small logarithmic term in (\ref{eq:w4}), we then obtain
\begin{equation}
\frac{b_{+,V}^{(6)}  + b_{S,V}^{(6)} }{ a_{+,V}^{(4)} + a_{S,V}^{(4)} } 
= \frac{1}{r_V^2}~.
\label{eq:ab6}
\end{equation}

In order to disentangle neutral
and charged channels beyond the above relation one has to rely on 
additional assumptions. We notice that out of the two 
solutions for $a_+$ and $b_+$ discussed in the previous section,
the one with $a_+ < 0$, consistent with
chiral counting, is also
compatible with the relation $b_+ / a_+ = 1/r_V^2$. 
This would follow directly from the assumption of 
vector meson dominance in the polynomial part of the form factor,
i.e. from the hypothesis
\begin{equation}
W_i^{\rm pol}(z)= a_i~\frac{r_V^2}{r_V^2 - z }~.
\label{eq:ab7}
\end{equation}

As commented previously, both form factors
$W_+^{\rm pol}$ and $W_S^{\rm pol}$ may receive short--distance type 
contributions that in principle could spoil this relation.
However, the form factor of the matrix element
$\langle \pi | \overline{s} \gamma^{\mu} d  |  K  \rangle$ 
exhibits a clear $K^*(892)$ pole dominance (in the $SU(3)$
limit it coincides with the electromagnetic form factor
$F(z)$ discussed in the previous section). Hence, we conclude 
that the pole structure of $W_i^{\rm pol}(z)$ 
is not restricted to the long--distance part.
Only the genuine ${\cal O}(p^6)$ weak vector transitions could 
spoil the relation (\ref{eq:ab7}). If these were negligible, 
and in general this is not necessarily the case, the ratio 
$b_i/a_i = 1/r_V^2$ would hold for both channels.

We can use this assumption to analyze
the $K_S \rightarrow \pi^0 \ell^+ \ell^-$ mode. Taking a look at the
expressions of $B(K_S \rightarrow \pi^0 \ell^+ \ell^-)$ in 
(\ref{eq:BRlist}) and considering that $a_S b_S > 0$ (as follows
from $b_S/a_S = 1/r_V^2$), we observe that for $|a_S|\gsim 0.2$
(i.e. for $B(K_S \rightarrow \pi^0 e^+ e^-)\gsim 2\times  10^{-10}$)
the constant and  linear terms in $a_S$ and $b_S$ are negligible and
can be dropped. In this way we obtain an expression for the branching 
ratios in terms of just one parameter
\begin{eqnarray}
B(K_S \rightarrow \pi^0 e^+ e^-) \, & \simeq & \, 5.2 \, a_S^2 \, 
\times 10^{-9}~, \nonumber \\
B(K_S \rightarrow \pi^0 \mu^+ \mu^-) \, & \simeq & \, 
1.2 \, a_S^2 \, \times 10^{-9}~,
\label{eq:brs}
\end{eqnarray}
from where we predict
\begin{equation}
\Frac{B(K_S \rightarrow \pi^0 \mu^+ \mu^-)}{B(K_S \rightarrow
\pi^0 e^+ e^-)} \simeq 0.23~.
\end{equation}
Unfortunately the lack of information on $a_S$ does not allow 
us to predict the separate rates. We note however that 
if $a_S \sim {\cal O}(1)$, as expected on general grounds, 
the electron mode should be within reach of the 
KLOE experiment \cite{DPHB}.

\section{$CP$ violation in $K \rightarrow \pi \ell^+ \ell^-$ decays}
\label{sect:CPV}

The dominance of single photon exchange 
does not apply to the $CP$--violating parts of 
$K \rightarrow \pi \ell^+ \ell^-$ amplitudes. 
In this case the hierarchy of the CKM matrix implies 
that $Z^0$--penguin and $W^\pm$--box diagrams, dominated by
short--distance contributions, play an important role.  
As a result, $CP$--violating amplitudes can be  
more conveniently studied by means of 
an appropriate four--fermion Hamiltonian 
that, in the case of $s\to d \ell^+\ell^-$ transitions, 
is known to next--to--leading order \cite{BL94}.
At scales $\mu < m_c$ it is given by
\beq
{\cal H}_{eff}^{|\Delta S| = 1} = \Frac{G_F}{\sqrt{2}} \, 
V_{us}^* V_{ud} \Big[ \, \sum_{i=1}^{6,7V} ( z_i(\mu) + \tau y_i(\mu) )
Q_i(\mu) ~+~ \tau y_{7A}(M_W) Q_{7A}(M_W) \, \Big]\, +\, \mbox{h.c.},
\label{eq:heff}
\eeq
where $\tau = - (V_{ts}^* V_{td})/(V_{us}^* V_{ud})$ and
$V_{ij}$ denote the CKM matrix elements. Here $Q_{1,2}$
are the current--current operators,  $Q_3,\dots,Q_6$ the QCD penguin
operators, whereas
\begin{equation}
Q_{7V} \, =  \, \overline{s} \gamma^{\mu}(1-\gamma_5) d \, \overline{\ell} 
\gamma_{\mu} \ell  \qquad \mbox{and}\qquad
Q_{7A} \, =  \, \overline{s} \gamma^{\mu} (1-\gamma_5) d \, \overline{\ell}
\gamma_{\mu} \gamma_5 \ell
\label{eq:q7v7a}
\end{equation}
are generated by electroweak penguin and box diagrams \cite{GW80,DDG}.
In the standard phase convention, the overall factor $V_{us}^* V_{ud}$
is chosen to be real and direct $CP$ violation is driven by 
the imaginary part of $\tau$. With this choice of phase,
only the QCD penguin operators and $Q_{7V},Q_{7A}$ are
relevant for estimating direct $CP$--violating amplitudes.
As can be understood by looking at the corresponding 
matrix elements, the dominant role
is played by $Q_{7V},Q_{7A}$  \cite{BL94}, with $y_{7V} (1~\mbox{GeV})
\simeq 5.7 \times 10^{-3}$ 
and $y_{7A}(M_W) \simeq -5.3 \times 10^{-3}$ \cite{BB96}
(corresponding to $\overline{m}_t (m_t) = 167 \,\mbox{GeV}$).

The contribution of $Q_{7V}$ to
$K \rightarrow \pi \ell^+ \ell^-$ decays 
interferes with the long--distance amplitude
discussed in the previous sections. In fact, the 
contribution proportional to the coefficient $z_{7V}$ 
is already included in the polynomial part
of the $K\to \pi\gamma^*$ form factor (as discussed in 
section \ref{sect:insights}). 
On the contrary, the interference of $Q_{7A}$ with the 
photon exchange amplitude vanishes as long as the lepton 
polarizations are summed over. 
In the following we shall discuss the role of $y_{7V}Q_{7V}$
in $K_L \rightarrow \pi^0 e^+ e^-$ and 
$K^\pm \to\pi^\pm e^+e^-$ decays.

\paragraph{4.1}
 From the definition (\ref{eq:tff}), $CP$ invariance would imply that 
the $K_L\to \pi^0 \gamma^*$ form
factor $W_L(z)$ vanishes. In the limit of 
$CP$ conservation, the decay $K_L \rightarrow \pi^0 e^+ e^-$ can only
proceed through the two--photon process  
$K_L \rightarrow \pi^0 \gamma \gamma  \rightarrow  \pi^0 e^+ e^-$
\cite{EPR88,CPC9,DG} or via subleading terms in the expansion of 
the $W^\pm$--box diagram \cite{BI}. 
The former mechanism is expected to be dominant yielding
$B(K_L \rightarrow \pi^0 e^+ e^-)_{CPC}\lsim ~\mbox{few}\times 10^{-12}$ 
\cite{DP96,CPC9,DG}. 

In the presence of $CP$ violation, the $K_L \rightarrow \pi^0 e^+ e^-$ 
transition can also proceed through
the small $\varepsilon_K K_1$ piece of the $K_L$ wave function
(indirect $CP$ violation) or via the direct $CP$--violating
amplitude generated by (\ref{eq:heff}). 
The contribution of $y_{7V}Q_{7V}$ appears in the function 
$W_L^{\rm pol}$, defined in analogy to (\ref{eq:Wpol}), with 
\beq
a_L \, = -\, i \Frac{4 \pi}{\sqrt{2}} \, \Frac{ \Im \lambda_t}{\alpha} \,
y_{7V} \qquad \mbox{and} \qquad \Frac{b_L}{a_L} \simeq \Frac{1}{r_V^2}~, 
\label{eq:albl}
\eeq
where $\lambda_t = V_{td} V_{ts}^*$.
On the other side, the indirect $CP$--violating contribution is given
in terms of
the parameter $\varepsilon_K$ as $A(K_L \rightarrow 
\pi^0 e^+ e^-)_{\rm ind}
= |\varepsilon_K| e^{i \pi/4} A(K_S \rightarrow \pi^0 e^+ e^-)$.
Summing the two vector--like $CP$--violating 
contributions, one has
\begin{equation}
W_L^{\rm pol} = 6.9 \times 10^{-4} \, \left[  3.3 \, a_S \, e^{i \frac{\pi}{4}} \, -\,  
i \Frac{ \Im \lambda_t}{10^{-4}}\right] \, \left[ 1 + \Frac{z}{r_V^2} \right]~.
\label{eq:wlpol}
\end{equation}
Unfortunately the sign of $a_S$ is not known. Moreover, 
we recall that for $|a_S| \gsim 0.2$ the dominant term in 
$B(K_S \rightarrow \pi^0 e^+ e^-)$ is provided by $a_S^2$,
therefore it is hopeless trying to extract information about the sign 
of $a_S$ from the measurement of 
$B(K_S \rightarrow \pi^0 e^+ e^-)$.\footnote{~In principle
the relative sign of $a_S$ with respect to $A(K_S\to\pi^+\pi^-\pi^0)$
could be measured by a careful analysis of the 
$K_S \rightarrow \pi^0 e^+ e^-$ spectrum (similarly to $a_+$ in
the charged mode). However, this program is made very difficult by
the smallness of  $A(K_S\to\pi^+\pi^-\pi^0)$. In addition, even if it were 
possible, one would then have to rely on a model--dependent 
assumption about the sign of  $A(K_S\to\pi^+\pi^-\pi^0)$ 
from (\protect\ref{eq:heff}).} 
However, since $a_S$ is real a strong interference 
(constructive or destructive) is expected
for $|a_S|$ and $|\Im \lambda_t/10^{-4}|$
of the same order of magnitude.

Including the $CP$--violating contribution of the $Q_{7A}$ operator that
does not interfere with (\ref{eq:wlpol}), we collect
all the $CP$--violating terms in $K_L \rightarrow \pi^0 e^+ e^-$ obtaining
\begin{equation}
B(K_L \rightarrow \pi^0 e^+ e^-)_{CPV} \, = \, 
\left[ 15.3\, a_S^2 \, - \, 6.8 \Frac{\Im \lambda_t}{10^{-4}} \, a_S \, + 
\, 2.8 \left( \Frac{\Im \lambda_t}{10^{-4}} \right)^2  \right] 
\times 10^{-12}~.
\label{eq:cpvtot}
\end{equation}
 From (\ref{eq:brs}) and (\ref{eq:cpvtot}) 
a very interesting scenario emerges for 
$a_S \lsim -0.5$ or    $a_S \gsim 1.0$.
Since $\Im \lambda_t$ is expected to be $\sim  10^{-4}$, one would 
have $B(K_L \rightarrow \pi^0 e^+ e^-)_{CPV}\gsim 10^{-11}$ in this case.
Then the $CP$--conserving contribution, which does not interfere with
the $CP$--violating part in the rate,
could be neglected. Moreover, the 
$K_S \rightarrow \pi^0 e^+ e^-$ branching ratio would 
be large enough to allow a direct determination of
$|a_S|$. Thus, from the interference term in 
(\ref{eq:cpvtot}) one could perform 
an independent  measurement of $\Im \lambda_t$,
with a precision increasing with the value of $|a_S|$.

A relation similar to (\ref{eq:cpvtot}) 
can be traced from the work of Dib et al. \cite{DDG} 
(see also \cite{BB96}). We note that this 
result arises from the VMD assumption about the $K_S\to\pi^0 e^+ e^-$ 
form factor discussed in section \ref{sect:insights}, which leads to
(\ref{eq:wlpol}). According to this 
hypothesis, 
the relative weight of direct and indirect $CP$--violating 
components in (\ref{eq:cpvtot}) is not affected by  
possible cuts in the $z$ spectrum. On the contrary, we recall that the 
$CP$--conserving rate does have a different (softer) spectrum
in $z$: essentially the 
factor $\lambda^{3/2}(1,z,r_\pi^2)$ in (\ref{fspectrum}) is 
replaced by $\lambda^{5/2}(1,z,r_\pi^2)$ \cite{EPR88,DG,BI}.
This fact could provide 
an additional handle for disentangling the various 
components of $K_L \rightarrow \pi^0 e^+ e^-$.

\paragraph{4.2}
The interference of the long--distance $K\to\pi\gamma^*$ amplitude 
and the short--distance contribution of $y_{7V}Q_{7V}$ 
leads to an asymmetry between the widths of $K^+ \rightarrow 
\pi^+ e^+ e^-$ and  $K^- \rightarrow \pi^- e^+ e^-$, which is a clear 
signal of direct  $CP$ violation \cite{EPR88}.
This quantity, defined by
\begin{eqnarray}
&& \Gamma (K^+ \rightarrow \pi^+ e^+ e^-) \, - \,    
\Gamma (K^- \rightarrow \pi^- e^+ e^-) \,  =  \, 
\Frac{G_F \, \alpha^2 \, M_K^3}{768 \, \pi^5} \,\times \nonumber \\
&&\qquad \int_{4 r_{e}^2}^{(1-r_{\pi})^2} \, 
dz  \, \lambda^{3/2}(1,z,r_{\pi}^2) \, \sqrt{1-4\Frac{r_e^2}{z}}
\left( 1+2 \Frac{r_e^2}{z} \right)\Im W^{\rm pol}_{+}(z)\, 
\Im W_{+}^{\pi \pi} (z)\, ~, \nonumber  
\end{eqnarray}
can be calculated unambiguously up to a sign. Indeed,
the only contribution to $\Im W_{+}^{\rm pol}$ comes from the 
short--distance regime and it is given by
\begin{equation}
\Im a_{+} \, = \, \Frac{4 \pi}{\sqrt{2}} \, \Frac{\Im \lambda_t}{\alpha}
\, y_{7V}~, \qquad\qquad\qquad 
\Frac{\Im b_{+}}{\Im a_{+}} \simeq \Frac{1}{r_V^2}~,
\label{eq:apbp}
\end{equation}
whereas the imaginary part of  $W_{+}^{\pi\pi}$
is determined by the physical 
$K^+\to\pi^+\pi^+\pi^-$ amplitude. Using the 
results in (\ref{eq:Wpp}-\ref{eq:cd})
and the experimental value of 
$\Gamma (K^+ \rightarrow \pi^+ e^+ e^-)$, we get
\begin{equation}
\Frac{|\Gamma(K^+ \rightarrow \pi^+ e^+ e^-) - \Gamma(K^- \rightarrow
\pi^- e^+ e^-)|}{\Gamma(K^+ \rightarrow \pi^+ e^+ e^-) + \Gamma(K^- 
\rightarrow \pi^- e^+ e^-)} \, = \, (0.071 \pm 0.007) \times
\left| \Im \lambda_t \right|~.
\label{eq:asr}
\end{equation}
The numerical value in (\ref{eq:asr})  
corresponds to an increase of a factor $\sim 2$ over the 
estimate at leading order in the chiral expansion \cite{EPR88}. However,
given that $\Im \lambda_t \sim 10^{-4}$, it is still
very difficult to detect such an effect, at least within the 
Standard Model.

\begin{figure}[t]
    \begin{center}
       \setlength{\unitlength}{1truecm}
       \begin{picture}(10.0,10.0)
       \epsfxsize 10.0 true cm
       \epsfysize 10.0 true cm
       \epsffile{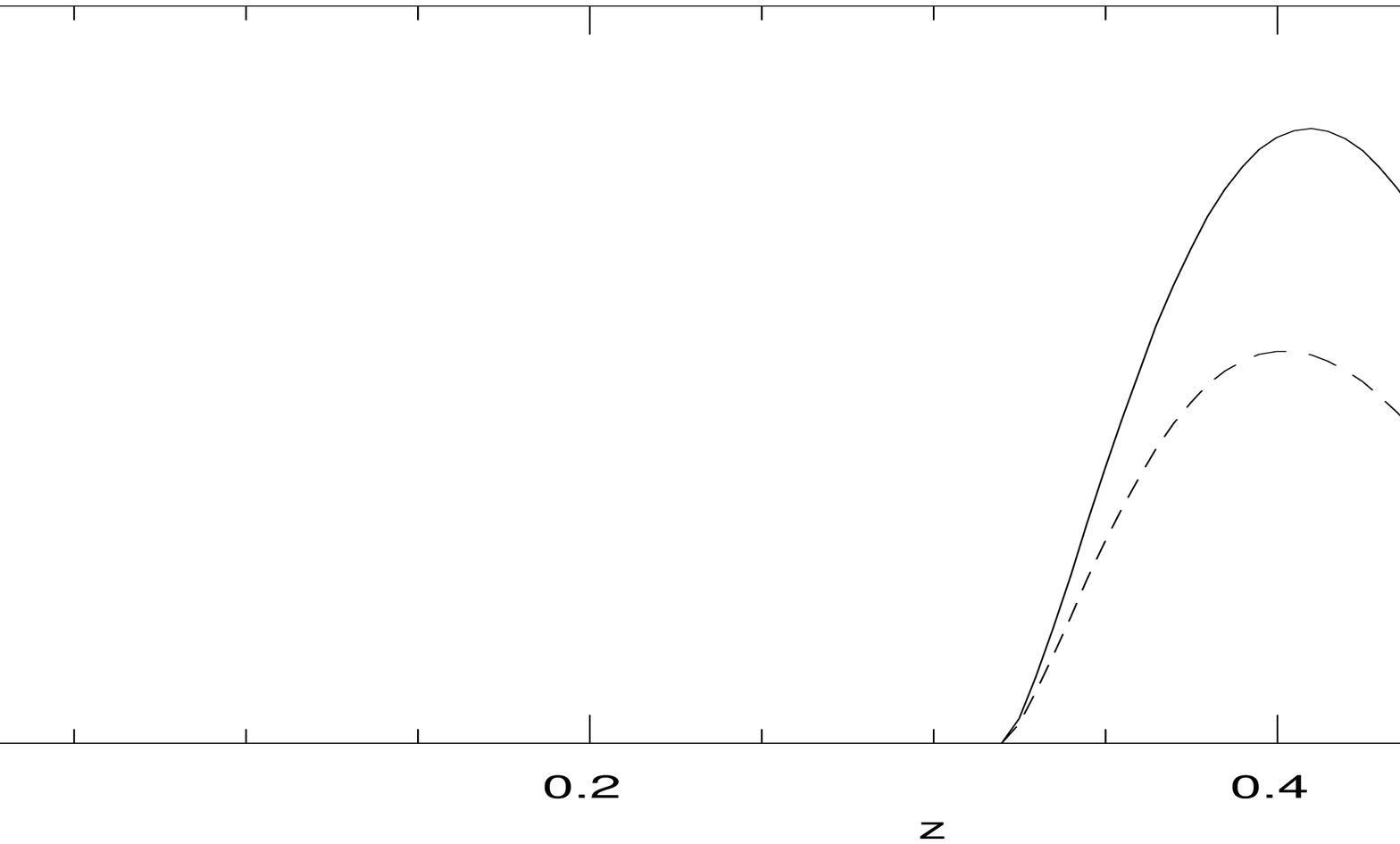}
       \end{picture} 
    \end{center}
    \caption{Differential charge asymmetry defined in
      (\protect\ref{eq:dgz}).
 The dashed line is the leading--order result in the 
 chiral expansion, whereas the full line includes the 
 unitarity corrections. } 
    \protect\label{fig:asyplot}
\end{figure}

A more interesting observable is the unintegrated asymmetry
defined by
\begin{equation}
\delta \Gamma (z) \, \equiv \, 
\Frac{\left| \Frac{d \Gamma}{dz}(K^+ \rightarrow \pi^+ 
e^+ e^-) \, - \, 
\Frac{d \Gamma}{dz}(K^- \rightarrow \pi^- e^+ e^-) \right| }{\Gamma (K^+
\rightarrow \pi^+ e^+ e^-) + \Gamma(K^- \rightarrow \pi^- e^+ e^-)}
\label{eq:dgz}
\end{equation}
that we have plotted in Fig.~\ref{fig:asyplot} both at leading 
order in the chiral expansion and including  unitarity
corrections. This asymmetry is seen to have a maximum 
around $z \simeq 0.4$ that, being far away from the dominant background,
could be more easily accessible 
from the experimental point of view than the integrated asymmetry.  

\section{Conclusions}

We have performed a model--independent analysis of 
$K^+ \rightarrow \pi^+ \ell^+ \ell^-$ and $K_S \rightarrow
\pi^0 \ell^+ \ell^-$ decays including  $K \rightarrow 3 \pi$
unitarity corrections and a general polynomial decomposition of 
the remaining dispersive amplitude. Using as input the presently 
available data on $K^+ \rightarrow \pi^+ e^+ e^-$ \cite{AL92}, 
we concluded that the 
ratio $R = B(K^+ \rightarrow \pi^+ \mu^+ \mu^-) / B(K^+ \rightarrow
\pi^+ e^+ e^-)$ must be larger than 0.23 unless there is a 
contribution from some non--standard physics. This result is to be
compared with the recent measurement $R = 0.167 \pm 0.036$
\cite{AD97}. 

We have shown that it is very difficult to predict 
$B(K_S \rightarrow \pi^0 e^+ e^-)$ without 
strong model--dependent assumptions, except for an approximate
upper bound $B(K_S \rightarrow \pi^0 e^+ e^-)$ $\lsim 10^{-8}$.
On the other hand, a realistic estimate 
of the muon/elec\-tron ratio can be obtained in the
framework of VMD: $B(K_S \rightarrow \pi^0 \mu^+ \mu^-) /
B(K_S \rightarrow \pi^0 e^+ e^-) \simeq 0.23$.

We have reanalysed the $CP$--violating contributions
to $K_L \rightarrow \pi^0 e^+ e^-$, emphasizing a possible strong 
interference (destructive or constructive) between direct and 
indirect $CP$--violating amplitudes. Together with a precise 
determination of $B(K_S \rightarrow \pi^0 e^+ e^-)$, this could 
provide a possible handle for isolating the direct $CP$--violating 
component. Finally, while the unitarity corrections lead to an 
increase of about $100 \%$ for the charge asymmetry
$\Gamma(K^+ \rightarrow \pi^+ e^+ e^-)$ --
$\Gamma(K^- \rightarrow \pi^- e^+ e^-)$,
the Standard Model prediction is still too small to be within reach 
of forthcoming experiments.

\vspace*{1cm}

\noindent{\bf Note added}
\vspace*{0.5cm} \\
At the recent ICHEP98 in Vancouver, preliminary results from a
high--statistics $K^+ \rightarrow \pi^+ e^+ e^-$ experiment (BNL-E865,
presented by Hong Ma) were announced. While the mean value of the
decay rate agrees with the previous measurement \cite{AL92}, the slope
is significantly bigger: $\lambda = 0.20\pm 0.02$. Although such a big
slope is still compatible with the theoretical expectation
$|b_+|<|a_+|$, the parameter $|b_+|$ would have to be bigger than the
naive VMD prediction $b_+=a_+/r_V^2$, but consistent with the 
factorization model predictions for the genuine  ${\cal O}(p^6)$ 
vector meson contribution.
Finally, we observe that the larger slope aggravates the discrepancy 
between theory and experiment for the $\mu/e$ ratio $R$.

\vspace*{1cm}
\noindent{\bf Acknowledgements}
\vspace*{0.5cm} \\
We thank F.J. Botella and  G. Buchalla for interesting discussions.
G.I. acknowledges the hospitality of the Theory Group at SLAC, where
this work has been completed. J.P. is partially supported by 
Grant AEN--96/1718 of CICYT (Spain).

\end{document}